# Do projects learn across space and time? Evidence from the Olympics

Atif Ansar[a,b], Bent Flyvbjerg[c,d,e] and Alexander Budzier[c]

[a]Saïd Business School, University of Oxford, Programme Director, Smith School of Enterprise and the Environment, University of Oxford, Oxford, UK; [b]School of Economics and Management, Tsinghua University School of Economics and Management, Beijing, China; [c]Saïd Business School, University of Oxford, Oxford, UK; [d]Villum Kann Rasmussen Professor and Chair of Major Program Management, IT University of Copenhagen, Copenhagen, Denmark; [e]St Anne's College, University of Oxford, Oxford, UK

**ABSTRACT**
Do projects learn across space and time? The Olympics, themselves among the largest publicly funded programmes in the world, offer a unique empirical setting. Theoretically, the Games seem ideal for generating 'positive learning curves', driving down costs from one iteration to the next. In practice, they do not. Drawing on the concept of 'myopia of learning', we argue that spatiotemporality (geographic distance, temporal gaps, and the temporary organizational form of each host committee) combines to block higher-level learning. Our analysis of cost overruns from 1960 to 2024 reveals no sustained improvement over 64 years. Tactical learning abounds, but none aggregates into strategic improvement. We propose four strategies for overcoming the spatiotemporal barrier (incremental, centralizing, decentralizing, and real options), arguing that radical reform is required.

## Introduction

Atlanta teemed with excitement as it hosted the 1996 Centennial Olympic Games. Behind the scenes, however, the city's transportation system was collapsing. Athletes, spectators, and officials faced delays and logistical nightmares due to insufficient infrastructure and traffic congestion. Amidst the chaos, British sculling gold medallists Steven Redgrave and Matthew Pinsent borrowed a car to reach their event at Lake Lanier. 'From our perspective, the organization here is diabolical', said Redgrave (Phillips 1996). Atlanta's transport problems were predictable. Pasqual Maragall, Mayor of Barcelona, which staged the 1992 Summer Games, commented, 'We did go through a painful four days also. It was a learning process' (Applebome 1996). Despite multiple, sequential opportunities to learn, the Rio 2016 Olympics faced similar transport issues with athletes stranded for hours and buses struggling to navigate Rio's complex, congested roads.

These lapses in planning and execution cascade into a deeper problem of direct relevance to public management: poor budget estimation and financial control in publicly funded projects.[1] Far from a well-oiled machine, the Olympics are among the worst-performing major project types when it comes to cost performance. With an average cost overrun of 159% in real terms, the Olympics perform worse than one-off US defence projects or NASA space missions (cf. Ansar 2023; Ansar and Flyvbjerg 2022; Flyvbjerg, Budzier, and Lunn 2021). Out of 23 project types studied, only nuclear waste storage projects have a higher average cost overrun (Flyvbjerg and Gardner 2023, 192). Given these escalating demands and chronic budget shortfalls, the viability of the Olympics in its current form is increasingly in question. This is not merely a sporting governance problem; it is a public management problem. Host cities commit vast public resources on the promise of legacy benefits, yet the evidence that these investments deliver value for money is, at best, mixed (Zimbalist 2016). Understanding why this pattern persists, and what might break it, has direct implications for how the public sector manages large-scale, spatially and temporally dispersed programmes.

The persistence of such problems from one Olympics to another is, at first, surprising. The Olympics seem to be an ideal setting for reinforcing learning from one Games to the next, a phenomenon we term positive learning curves. After all, the Games exemplify repeatable programmes, occurring every four years. The type and number of events are largely consistent and should therefore be ideal for effective learning. Why, then, is delivering repeatably successful Olympics so challenging, and so rare?[2] We argue that the answer lies in the *spatiotemporal dispersion* of the Games – the fact that each edition is staged in a different city, by a different organization, separated by intervals of years. Drawing on Levinthal and March's (1993) concept of the 'myopia of learning', which identifies how organizations become trapped in spatial, temporal, and failure-related shortsightedness, we show that the Olympics' rotational hosting format systematically blocks the accumulation and transfer of higher-level organizational knowledge. Where Levinthal and March theorized myopia in general terms, we develop its spatial and temporal dimensions into the concept of *spatiotemporality* (the compound effect of geographic and temporal distance on an organization's capacity to learn) and apply it to the empirical setting of the Olympic Games.

A note on what we mean by 'the Olympics' is essential here, because the unit of analysis matters for how we understand learning. In this paper, 'the Olympics' refers to the institutional ecosystem responsible for staging the Games. This ecosystem operates as a *franchise* (cf. Gillett and Tennent 2017, 109). The International Olympic Committee (IOC) functions as the permanent franchisor: it owns the brand, sets the rules, and coordinates knowledge transfer.[3] But the IOC does not deliver the Games. Each edition is delivered by a bespoke, temporary organization – the Organising Committee for the Olympic Games (OCOG) – which functions as the franchisee, assembling local and national stakeholders to execute a one-off mega-event before disbanding. International Federations (IFs) define the technical requirements for each sport; National Olympic Committees (NOCs) manage national participation. This loose, federated structure means that the Olympics are not one organization learning repeatedly, but a series of *temporary organizations* (Lundin and Söderholm 1995) operating under the broad oversight of a permanent institution that lacks direct operational control. The distinction matters: learning in temporary organizations is inherently constrained by their predefined life cycles, transient teams, and task-specific mandates (Beck et al. 2024; Marrewijk, Stjerne, and Sydow 2024). It is precisely this organizational form, not merely the content of the Games, that produces the learning failures we document.

In this paper, we make three contributions. First, we present the most comprehensive statistical analysis of Olympic cost performance to date, covering all available Games from 1960 to 2024, and show that there has been no sustained reduction in cost overruns, their magnitude, frequency, or variance over 64 years. This evidence anchors our central claim: the Olympics have not achieved higher-level learning across time and geography. Second, we develop the concept of spatiotemporality as a framework for understanding why dispersed project ecologies (Osborne et al. 2022) fail to learn, extending Levinthal and March's (1993) myopia of learning into the spatial and temporal domains. We distinguish between lower-level learning – tactical improvements within existing routines, of which the Olympics show abundant qualitative evidence – and higher-level learning – strategic shifts in organizational norms and delivery models, of which the cost data reveal none. Third, we propose four strategies for addressing this learning deficit: incremental, centralizing, decentralizing, and real options, arguing that the Games' reliance on incremental approaches has been insufficient and that more radical structural reform is required.

Organizational learning in public management is critical yet under-researched (Rashman, Withers, and Hartley 2009). There is a particular need for what Bouckaert and Peters (Bouckaert and Guy Peters 2002, 360) call a 'helicopter view' (what we term higher-level learning) to overcome the more typical myopic focus on immediate operational fixes. The Olympics illustrate this challenge vividly: public institutions routinely commit billions in taxpayer funds to mega-events based on optimistic forecasts, only to confront cost overruns that erode public value. Such challenges are not unique to the Games (see Cinar, Trott, and Simms 2019, on the systemic and process-wide barriers to public sector innovation). Public administration in large federated structures – the U.N., the World Bank, and national systems like those of India, Brazil, and Australia – faces analogous problems of learning across spatial and temporal distance.[4] Recent work in this journal underscores related themes: the fragility of learning in temporally dispersed peacekeeping settings (De Waard et al. 2023), the tensions of cross-boundary collaboration (Brorström and Diedrich 2022), the importance of broad knowledge search (Burgers, Arundel, and Casali 2024), and the challenges of network governance in Olympic stakeholder ecosystems (dos Santos, Monteiro, and Saad 2025). Rizzo et al. (2025) show how ecosystem design shapes public value creation in the Milano-Cortina 2026 Winter Games, while Barría Traverso (2025) demonstrates how a country's political system conditions the value framework within which a mega-event takes place. Our analysis contributes to these conversations by offering a longitudinal, empirically grounded account of why spatiotemporal dispersion systematically undermines learning in public project delivery.

The paper proceeds as follows. We first develop our theoretical framework, drawing on the myopia of learning, the distinction between higher- and lower-level learning, and the impacts of spatial and temporal distance on project performance, advancing four propositions. We then present the governance context of the Olympic Games before turning to our results: a statistical analysis of cost overrun data from 1960 to 2024, followed by qualitative evidence of learning dynamics. The discussion develops our concept of spatiotemporality, evaluates the four proposed strategies, and situates our findings within the broader literature on public management and project learning. We conclude with implications for theory and practice.

## Theoretical background: in search for organizational intelligence

As outlined in the Introduction, projects like the Olympics are temporary organizations, which shapes how they acquire and retain knowledge. Throughout this paper, we define projects as an organizational form – temporary, task-specific structures with finite life cycles (Lundin and Söderholm 1995) – rather than as a process or set of management techniques. Building on this foundation, this section examines in more depth the theoretical perspectives most relevant to our analysis, particularly processual views of project success and failure, and the concept of Temporal Adaptive

Capacity (TAC) as applied to successive Olympic Games (see, e.g. Beck et al. 2024; Lundin and Söderholm 1995; Marrewijk, Stjerne, and Sydow 2024).

The inherent difficulty of learning across temporary organizations stems from low fidelity in knowledge transfer, mismatched organizational structures, wilful ignorance, and organizational forgetting (De and Phillips 2012; Easterby-Smith and Lyles 2011; Grabher 2004; Gross and McGoey 2015; Levinthal and March 1993; Levitt and March 1988; Swan, Scarbrough, and Newell 2010; Thiel and Grabher 2024). These challenges are magnified in mega-event ecologies, where multiple layers of temporary and permanent organizations interact across different time scales. In the case of the Olympic Games, learning is further complicated by:

- Rotational Host City Organizations: Responsible for frontline delivery yet operating on a single-cycle basis, limiting institutional memory.
- The International Olympic Committee (IOC): Tasked with ensuring continuity between Games but often struggles to retain and transfer knowledge effectively.
- Diverse, Multi-Stakeholder Ecosystems: Public institutions, private enterprises, and hybrid entities each bring competing priorities, governance structures, and incentives, making coordinated learning difficult.

To better understand these challenges, Gillett and Tennent (2017) analyse the 1966 FIFA World Cup in England, applying two complementary mega-project frameworks: Flyvbjerg's four sublimes (technological, political, economic, and aesthetic) and Morris and Geraldi's three levels of project management (technical, strategic, and institutional). Their analysis highlights how host organizations' motives evolve over time, leading to shifting project goals. Similar to the Olympics, the 1966 World Cup required extensive multi-stakeholder coordination, involving government agencies, sports federations, local organizing committees, and host venues. Their study reveals those fluctuating objectives, whether economic stimulus, national prestige, or sporting legacy, complicate governance and budgetary discipline.

Gillett and Tennent's (2017) 'dynamic sublimes' framework underscores that learning in mega-event ecologies is rarely straightforward. When host cities or national governments alter project scope midstream, by injecting public funds, revising regulations, or introducing new requirements, earlier plans are frequently overwritten, leading to short-term tactical shifts and long-term unintended consequences. This mirrors the recurring challenges in the Olympics, where cost overruns and fractured learning persist across different Games. Their study further underscores that 'top-level' organizations like FIFA and the IOC seldom act alone; instead, they operate within diffuse governance structures, where local clubs, city councils, and national ministries significantly shape final outcomes. This aligns with broader research on Olympic governance, which demonstrates that fragmented authority and misaligned incentives hinder knowledge transfer between Games.

Despite efforts at incremental learning, such as stadium improvements, new broadcast technologies, and enhanced logistical planning, political incentives often overshadow purely operational lessons (Gillett and Tennent 2022). This finding resonates with research on Olympic host cities, where replicating best practices remains challenging due to local contextual variability and inconsistent knowledge retention mechanisms. The 1966 World Cup case reinforces that megaproject learning is constrained by continual shifts in stakeholder composition and project objectives, further supporting the argument that prestige, economic interests, and political agendas often take precedence over systematic learning and cost containment.

By situating mega-event ecologies within the broader theory of temporary organizations, this discussion highlights how shifting stakeholder motives, governance complexity, and the interplay of temporary and permanent organizations impede the institutionalization of knowledge. As knowledge transfer remains inconsistent and politically influenced, host cities and governing bodies often prioritize short-term problem-solving over long-term institutional learning. This tendency to focus on immediate execution rather than cumulative improvement reflects a deeper pattern of organizational myopia, where lessons from past events are either selectively remembered or entirely forgotten, shaping how future projects unfold.

## *Myopia of learning*

Levinthal and March (1993) advance the concept 'myopia of learning'. They delve into the limitations of learning processes within organizations, particularly focusing on the behavioural concept of myopia, or shortsightedness, in organizational learning. The authors argue that while learning is crucial for the adaptive capabilities of organizations, it inherently involves certain biases and limitations that can adversely affect long-term performance and adaptability.

They identify three forms of myopia in organizational learning: spatial, temporal, and failure. Spatial myopia refers to the limitation of organizational learning to local or nearby contexts, where organizations tend to overlook lessons and feedback that could be gleaned from more distant or disparate areas. Temporal myopia occurs when organizations prioritize short-term gains and immediate benefits over long-term stability and future opportunities, often neglecting the long-term sustainability of their actions. Moreover, organizations often overly anchor on recent data, ignoring longer-term historical data relevant to forming a debiased view. This narrow focus can prevent them from recognizing broader trends and variations that might affect their strategic position.

Failure myopia in organizational learning, as described by Levinthal and March, refers to the tendency of organizations to focus disproportionately on successes while neglecting or undervaluing failures. This bias leads to a skewed perception where successful outcomes are overrepresented in the learning process, resulting in an organizational history that underrepresents failures (see Shepherd, Patzelt, and Wolfe 2011). Such myopia can cause organizations to become overconfident in their abilities and strategies, underestimating the risks and overestimating their control over outcomes. This often results in inadequate preparation for dealing with future challenges and reduces the organization's ability to learn from past mistakes, potentially stifling performance and adaptive growth.

Levinthal and March are cautious about whether organizations attempting to manage and mitigate these learning myopias can be successful. While organizations strive to simplify and specialize to enhance learning effectiveness, these very actions can embed myopic tendencies deeper within the organizational processes. The authors conclude by suggesting that although these learning imperfections are significant, they do not necessarily warrant abandoning efforts to enhance organizational learning capabilities. Rather, they advocate for a tempered approach with moderated expectations regarding the outcomes of organizational learning initiatives.

Considering Levinthal and March's (1993) caution, supported by a broader empirical literature, we advance the following:

**Proposition 1:** *Organizational learning does not spontaneously arise from routine operations; it demands strategic design for its effective realization.*

### *Higher and lower-level learning*

In their influential work, Fiol and Marjorie (1985) support the proposition that organizational learning is a deliberate process, not an incidental outcome of mere time passing. The essence of their argument is that while organizations inherently adjust to internal pressures and external environmental changes over time, these adjustments should not be confused with genuine learning. True learning implies a substantive improvement in organizational actions rooted in enriched knowledge and a deeper understanding, which extends beyond mere adaptations or tactical responses to immediate challenges.

The nuance between adaptation and learning is salient. Fiol and Lyles distinguish between different levels of learning. Lower-level learning is characterized by short-term, superficial changes that often reinforce existing behaviour patterns without substantial cognitive development. For example, an organization might tighten its invoicing procedures after discovering an incidence of petty fraud to ensure future suppliers are less likely to overcharge. Higher-level learning, however, involves profound shifts in organizational norms, leading to long-term strategic changes. This learning involves deep cognitive processes such as reflection, re-evaluation, and the restructuring of organizational norms and worldviews. Higher-level learning leads to the modification of the fundamental assumptions underpinning organizational strategies and may catalyse comprehensive changes in organizational direction and identity. For example, Netflix's ability to transform itself from a DVD rental business to an online streaming platform and then to an original content creator qualifies as higher-level learning (Hastings and Meyer 2020; Jaworski 2021).

These levels of learning resonate with Levinthal and March's concept of the myopia of learning, which highlights how organizations often become trapped in short-term thinking, failing to explore new knowledge that might benefit them in the longer term. Senge's ([1990] 2006) concept of the 'learning organization' – in which systems thinking, shared mental models, and team learning combine to produce adaptive capacity – captures what higher-level learning looks like when it works. Yet Senge's framework presupposes organizational continuity: stable teams, persistent institutional memory, and iterative feedback loops. These are precisely the conditions that the Olympics' temporary, rotating structure denies. The higher and lower levels of learning also help us understand why participants inside an organization may misconstrue operational experience as learning (lower-level learning) even when higher-level outcomes do not improve. In formal terms:

**Proposition 2:** *Learning is unevenly distributed inside organizations. An organization may enhance its routine procedures (lower-level learning) yet still not achieve strategic success (high-level outcomes).*

### *Spatiotemporality and the impacts of space and time on learning*

Levinthal and March (1993) treat spatial and temporal myopia as analytically distinct. In practice, they compound. We use the term *spatiotemporality* to capture this interaction – the way geographic and temporal distance combine to erode an organization's capacity for higher-level learning. Spatial distance degrades tacit knowledge transfer (Hansen and Nohria 2004); temporal distance ensures that even codified knowledge decays before the next iteration (Argote, McEvily, and Reagans 2003; Walsh and Ungson 1991). When both operate simultaneously, the barriers are not additive but multiplicative. The effect is amplified by organizational form: in permanent organizations, stable teams and institutional memory can mitigate spatiotemporal challenges, but in temporary organizations, such as the OCOGs that deliver each Olympic Games, these structures are absent by design (Beck et al. 2024; Lundin and Söderholm 1995). The Olympic franchise model, in which a permanent franchisor (the IOC) delegates delivery to a rotating series of temporary franchisees (the OCOGs), represents an extreme case: maximum geographic distance combined with maximum organizational discontinuity.

Spatiotemporality also connects to recent work on project ecologies and public value. Osborne et al. (2022) argue that value in public services emerges from interactions within ecosystems shaped by proximity, timing, and relational architecture. Rizzo et al. (2025) demonstrate this in the Milano-Cortina 2026 Winter Games, showing how ecosystem design conditions public value creation, Geeraert and Van Bottenburg (2025) show how the pursuit of public value can sustain collaborative governance in sport mega-events, while dos Santos, Monteiro, and Saad (2025) reveal that 'local' Olympic actors, closest in space but most constrained in time, are often marginalized in knowledge flows. Spatiotemporality is thus not merely a description of the Olympic format but a structural condition shaping what kinds of learning are possible. We now consider the separate and combined impacts of space and time on learning across projects.

### *Space*

'Heaven is high, and the Emperor far away'— Ancient Chinese proverb[5]

Spatially dispersed projects may have worse performance than projects happening in a single location due to a variety of factors related to culture and communication, coordination, and knowledge sharing that arise from spatial dispersion.

Geographically dispersed teams face language barriers, cultural differences, and more prosaic problems such as time zone differences, which often lead to communication problems. Hinds and Mortensen (2005) found that spatial dispersion negatively impacts shared understanding and communication effectiveness in teams. Similarly, building on a vast literature in cultural anthropology, Meyer (2014) demonstrates that cultural misunderstandings are all too likely in multicultural teams unless teams take deliberate steps to understand and adapt to various cultural dimensions. According to Meyer's research, business cultures differ across eight dimensions: communication styles (high-context vs. low-context), feedback approaches (direct vs. indirect), persuasion strategies (results-first vs. methods-first), leadership preferences (egalitarian vs. hierarchical), decision-making processes (consensual vs. top-down), trust-building methods (task-based vs. relationship-based), disagreement styles (confrontational vs. conflict-averse), and time management orientations (linear-time vs. flexible-time).

The spatial and cultural distance creates coordination and knowledge transfer issues. Geographically dispersed projects often involve coordinating work across multiple locations and time zones. This can result in coordination challenges and delays, as discussed by Cummings and Haas (2012), who observed that coordinating interdependent tasks is more difficult in spatially dispersed teams. The geographic separation of team members can make it more difficult for them to share and access tacit knowledge, which is essential for innovation and problem-solving. Hansen and Nohria (2004) found that distance reduces the likelihood of knowledge sharing, particularly when it comes to tacit knowledge. The spatial barriers erode social capital and trust. Jarvenpaa and Leidner (1999) observed that building trust is more challenging in virtual teams, while Mortensen and Hinds (2001) found that spatially dispersed teams experience a lower level of trust and shared identity compared to co-located teams.

In contrast to spatial dispersion, geographically concentrated projects benefit not only from the traditional economies of scale but also from agglomeration economies, such as knowledge spillovers, access to specialized suppliers, and a skilled labour pool. Spatial proximity enables the rapid exchange of information, ideas, and innovation between firms and individuals (Jaffe, Trajtenberg, and Henderson 1993; Storper and Venables 2004). Jaffe, Trajtenberg, and Henderson (1993) find that knowledge spillovers, as indicated by patent citations, are more likely to occur within geographically proximate areas. Moreover, agglomeration facilitates access to specialized suppliers and institutions, as highlighted by Porter (1998) in his work on industry clusters. Geographically concentrated projects also benefit from a skilled labour pool, as the density of economic activities in cities attracts talent and promotes specialization (Glaeser et al. 1992). This concentration of skilled workers and specialized suppliers allows for the creation of regional advantages, as demonstrated by Saxenian (1996) in her research on Silicon Valley and Route 128. In these technology hubs, regional culture and social networks play a crucial role in facilitating collaboration, knowledge sharing, and access to resources, resulting in increased entrepreneurial activity and economic growth. The benefits of agglomeration economies are not as easily accessible to dispersed projects (Fujita, Krugman, and Mori 1999). Even with the increasing prevalence of work-from-home norms, Bratton and Wójcik (2024) found that the lack of acute physical proximity interrupted knowledge transfer for equity research analysts.

Pankaj Ghemawat's work is seminal in organizing these various facets of spatial distance into a coherent framework. Ghemawat (2007) illustrated that there are four dimensions of distance: Cultural, Administrative, Geographic, and Economic, which together form the CAGE acronym. Administrative distance refers to differences in political, legal, and institutional environments between countries. In the context of geographically dispersed projects, the type of law (common vs civil), variations in regulations, labour laws, and taxation can create additional complexities and challenges for project management, requiring a deeper understanding of local contexts. Geographic distance encompasses not only physical distance but also factors such as time zones, transportation infrastructure, and climate. Economic distance involves differences in income levels, resource endowments, and market structures (Ghemawat 2007).

These differences can affect the availability of resources, labour costs, and market access, which may have significant implications for the cost, quality, and overall success of geographically dispersed projects. In formal terms:

**Proposition 3:** *As cultural, administrative, geographic, and economic distances among projects increase, the overall performance will tend to deteriorate and exhibit greater variability.*

### Time

*'The past is a foreign country: they do things differently there'.* — L. P. Hartley

Temporally dispersed projects are at a disadvantage compared to projects happening more regularly due to factors related to learning curves, organizational memory, resource allocation, and team cohesion. One of the most enduring concepts related to performance over time is the learning curve, which theorizes that as an organization repeats a task, its efficiency and effectiveness improve over time (Wright 1936). T.P. Wright's seminal paper demonstrated that as the production of airplanes increased, the labour input per unit decreased at a predictable rate. In contrast, for temporally dispersed projects, infrequent occurrences hinder the opportunities for organizations to learn from their experiences and refine their processes, resulting in poorer performance compared to more regular projects. A substantial body of work on learning curves underpins these conclusions (see, for example, Adler and Clark 1991; Argote and Epple 1990; Lapré and Nembhard 2010; Yelle 1979).

The learning curve is inherently linked to time, as it captures the relationship between the cumulative experience or output of an organization and the improvements in efficiency, effectiveness, cost reduction or increase in speed-to-market over time (see Ansar and Flyvbjerg 2022). The literature discusses the link between the learning curve and time through several factors such as learning by doing (Arrow 1962); learning from mistakes (Argote and Epple 1990); accumulation of knowledge and its overall impact in increasing the 'absorptive capacity' of organizations (Cohen and Levinthal 1990); organizational routines such as standardized procedures that streamline processes that reduce variability (Nelson and Winter 1982); and technological advancements. With time, organizations may adopt new technologies or improve existing ones, leading to increased efficiency and improved performance.

Technological advancements can also facilitate knowledge sharing and collaboration, further accelerating the learning process (Zollo and Winter 2002).

Time also erodes organizational memory, which is the stored knowledge within an organization crucial for its efficient functioning (Walsh and Ungson 1991). Temporally dispersed projects often suffer from memory decay, where knowledge from previous projects is forgotten, misremembered, or outdated, hindering knowledge transfer and performance (Cross and Baird 2000; Davenport and Prusak 1998; Levitt and March 1988). Factors like personnel turnover, retirement, and team reconfiguration exacerbate this issue (Argote, McEvily, and Reagans 2003; Darr, Argote, and Epple 1995; Szulanski 1996). Team familiarity and cohesion positively impact project performance (Reagans, Argote, and Brooks 2005). Teams with prior collaboration develop 'deep-level diversity' through meaningful interactions (Harrison, Price, and Bell 1998). Temporally dispersed projects struggle to maintain team cohesion due to less experience working together and reassignment during gaps. These projects also compete for resources with regular operations (Swan, Scarbrough, and Newell 2010). Similarly, the Olympic bidding process constrains learning: because host cities only confirm their status after securing the Games, the time available to systematically acquire and adapt lessons from previous hosts is compressed, often committing them to sub-optimal knowledge search and transfer.

Phase transitions, such as transitioning from design to construction, result in significant information loss. For instance, the phaseout of Germany's nuclear power plants poses challenges due to the incomplete or missing historical design information that is crucial for decommissioning (Scherwath, Wealer, and Mendelevitch 2020).

Organizational memory decay can occur quickly, especially during handovers between project teams, like from development to execution, losing valuable decision-making information (Dyck et al. 2002). Insufficient information gathering and slow decision-making further accelerate memory loss (Eisenhardt 1989); so does technological obsolescence. For all these reasons, organizations face 'catastrophic forgetfulness'[6] (French 1999; Kirkpatrick et al. 2017; McCloskey and Cohen 1989; Zenke, Poole, and Ganguli 2017). The CEO of a large real estate investment firm said to one of us in an interview, 'We've been constructing projects since 1973. Yet every time we start a new project, it feels as though we've never built anything before'. From this discussion, we derive the following:

**Proposition 4:** *Projects with temporal dispersion will tend to perform worse than more frequent projects, with mediating factors including inadequate record-keeping, information loss, and alterations in team composition.*

### *Spatiotemporality: the compound challenge*

Having considered spatial and temporal dispersion separately, we can now characterize their interaction. Spatial distance ensures that knowledge is context-dependent: what works in Barcelona may not transfer to Beijing, because regulatory environments, labour markets, and political incentives differ across every dimension of Ghemawat's (2007) CAGE framework. Temporal distance ensures that even transferable knowledge degrades before it is needed (Dyck et al. 2002; Walsh and Ungson 1991). In the Olympic Games, where each edition is staged in a new city by a new organization after a four-year interval, the barriers are multiplicative: the new host must adapt knowledge from a distant context using fragmentary information transmitted by an organization that no longer exists.

The project management literature has increasingly recognized learning as a central concern (see, e.g. Brady and Davies 2004; Hartmann and Dorée 2015; Prencipe and Tell 2001), with recent work examining the role of project memory in large-scale settings (Mariano and Awazu 2024), the mechanisms through which individual actors transfer knowledge across strategic projects (Eikelenboom and van Marrewijk 2024), and the deliberate facilitation required to embed learning within project routines (Dowson, Unterhitzenberger, and Bryde 2024). Much of this work, however, focuses on settings where spatiotemporal distance is limited – projects by a single organization in one location, or spatially dispersed projects managed by a permanent firm. The Olympics represent an extreme case in which neither condition holds, making them a critical case (Flyvbjerg 2006) for testing theories of project learning at their limits. Barría Traverso (2025) reinforces this point, showing that the political system of the host country further conditions the severity of the spatiotemporal challenge: the greater the CAGE distance between successive hosts, and the longer the interval between them, the higher the barrier to learning.

The compound effect of spatiotemporality is to dilute the overall efficacy of learning pathways. In other words, it is difficult for an organization to keep it together in one place at one time. When both the space and time are liquidated, something extraordinary is required to transfer learning. As we see in the next section, the Olympics fail to meet the high bar.

### Research design and data

This study adopts a longitudinal, historical research design to examine whether organizational learning occurs across successive Olympic Games. Following Tennent and Gillett (2024), we treat historical project data not merely as a descriptive record but as an empirical basis for developing and testing theory, in our case the propositions derived from Levinthal and March's (1993) myopia of learning and our concept of spatiotemporality. The research design combines quantitative analysis of cost overrun data with qualitative evidence drawn from the secondary literature, deploying each for a distinct analytical purpose.

### Quantitative data and analysis

The quantitative component draws on a dataset of Olympic Games cost overruns originally compiled by Flyvbjerg, Budzier, and Lunn (2021) and updated by Budzier and Flyvbjerg (2024). The dataset covers all Games from 1960 to 2024 for which reliable cost data are available: 23 out of a total of 34 Games. Cost overruns are measured as the percentage difference between actual outturn sports-related costs and the cost estimate at the time of bidding, in real terms (i.e. excluding inflation). The dataset is the largest and most consistent of its kind; for full details of data collection and measurement, see Flyvbjerg, Budzier, and Lunn (2021) and Budzier and Flyvbjerg (2024).

We assess four empirical markers of higher-level learning: the frequency of cost overruns, their magnitude (mean overrun), their variability (standard deviation), and the occurrence of radical changes in delivery format. These markers are grounded in the organizational learning literature, where sustained reductions in error frequency, magnitude, and variance are established indicators that learning is taking place (Argyris and Schön 1978; Weick and Sutcliffe 2011). We apply a suite of statistical methods (linear and polynomial regression, LOESS smoothing, and ARIMA(0,0,4) time-series modelling) to test for trends over time. Mann-Whitney U tests are used to compare cost overrun distributions across regions and between Summer and Winter Games.

A limitation of the quantitative data is its incompleteness: for 11 of 34 Games since 1960, valid and reliable cost data could not be found. Coverage of Summer Games is particularly thin prior to the 1990s, with no data available between Montreal 1976 and Barcelona 1992. This constrains statistical power and means that trend estimates for earlier decades should be interpreted with caution. Additionally, two data points (Beijing 2008 and PyeongChang 2018, each showing 2% overruns) may reflect reporting conventions rather than genuine cost discipline, although per-athlete expenditures are consistent with norms for their respective Games types. We retain these figures, noting that any understatement would lower, not raise, the aggregate estimate.

### Qualitative data and analysis

The qualitative component draws on the published academic literature, official IOC reports, and contemporaneous media accounts to document instances of learning within and across Olympic Games. This evidence is used to assess lower-level learning (tactical improvements in areas such as crowd management, venue accessibility, security, anti-doping, and transport coordination) which is not captured by the cost overrun data.

### *Rationale for the mixed approach*

The combination of quantitative and qualitative evidence is deliberate and serves a specific analytical function. Cost overrun data provide a consistent, comparable metric for assessing higher-level learning, that is, whether the Olympic ecosystem has achieved sustained improvements in strategic outcomes over time. Qualitative evidence provides a complementary lens for identifying lower-level learning: tactical adaptations that occur within individual Games but may not aggregate into strategic improvement. Together, the two forms of evidence allow us to distinguish between these levels of learning, which is central to our theoretical argument.

This approach has strengths: the cost data offer a longitudinal, cross-Games comparison that is rare in the mega-event literature, while the qualitative evidence grounds the analysis in the operational realities of Olympic delivery. It also has limitations. Cost overruns are a proxy for higher-level learning, not a comprehensive measure of it; they do not capture improvements in non-financial dimensions such as athlete experience, broadcast quality, or legacy outcomes. The qualitative evidence, drawn from secondary sources, varies in depth and coverage across Games and is inevitably shaped by the availability and accessibility of published accounts. We do not claim that these data capture all dimensions of organizational learning; rather, they provide one critical lens that must be interpreted in combination.

### **Are the Olympics learning? Results**

Learning across projects in organizations is challenging, but the empirical signs of whether an organization or a project ecology is learning are hard to miss. The literature proposes three robust empirical markers of whether learning is taking place: the frequency, magnitude, and variance of errors in a process. Argyris and Schön (1978) conceptualized this through the paradigms of single-loop and double-loop learning, where the former addresses immediate errors within existing organizational routines and the latter challenges and refines the underlying assumptions leading to those errors. This framework is particularly relevant in project management contexts, where learning is evidenced by the reduction in cost overruns and variance across projects over time.

The theory of High Reliability Organizations (HROs) further illustrates this by examining entities that operate in environments where the cost of error is exceptionally high, such as nuclear reactors or hospitals, and maintain low error rates through robust organizational cultures and error-checking processes (Weick and Sutcliffe 2011). Additionally, the Organizational Information Processing Theory suggests that enhancing how information is processed within an organization can significantly reduce errors and stabilize outputs (Galbraith 1974). In practice, this often translates into the adoption of tools like Statistical Process Control (SPC) and Root Cause Analysis (RCA), which help in identifying and correcting deviations from desired performance standards. For example, Six Sigma is a statistical term used to describe the maximum number of defects a process, generalizable across fields such as manufacturing, FedEx parcel deliveries, pathology test, or product cost over-runs, must produce to meet the Six Sigma standard, which is only 3.4 errors per one million units of output (Choo, Linderman, and Schroeder 2007; Eckes 2002; Schroeder et al. 2008; Tosuner et al. 2016). Organizations that effectively integrate structured data analysis and a culture of continuous feedback are better positioned to demonstrate learning through improved project cost metrics and reduced variability in project outcomes (Senge [1990] 2006).

Applying this literature to project management, in a previous comparative study of NASA vs SpaceX, we showed that the data markers of learning across projects include: reduction in the frequency and magnitude of cost and time overruns, reduction of variance in the cost and time overruns, lower cost-of-production, faster speed-to-market, and great scalability (see Ansar and Flyvbjerg 2022). Our reference class dataset consisted of 203 space missions spanning the years 1963–2021, comprising 181 missions from NASA and 22 from SpaceX. We found that SpaceX's project-delivery strategy was 10X cheaper and 2X faster than NASA's bespoke strategy. Moreover, SpaceX's platform strategy was less risky, virtually eliminating cost overruns. We further showed that SpaceX is more scalable – in 2023, SpaceX launched a total of 98 space flights compared to fewer than 10 for NASA.

### Higher-level learning at the Olympics

Let us now turn to the data on the Olympics and analyse the frequency, magnitude, and variance of cost estimation errors, using the dataset and methods described above. The full dataset is provided in the Appendix, including Table A1.

Our assessment of the Olympic Games focused on four key learning markers: the frequency of cost overruns, the magnitude of these deviations (measured as the mean cost overruns), the variability in the magnitude of errors (measured as the standard deviation of overruns), and the occurrence of radical changes in delivery format. The data indicate significant cost overruns and high variability, suggesting ineffective learning (see Table 1).

Table 2 shows cost overruns for the Olympics across various regions. Pairwise Mann-Whitney U tests were used to compare the distributions of cost overruns between each pair of regions. The results revealed no

Table 1. Magnitude, frequency, and variability of cost errors at the Olympics, cost overrun in real terms.

|  | Olympics Actual Performance 1960–2024 | Theoretical Expectation: If the Olympics Were Learning* | Practical Example of Learning: SpaceX** |
|---|---|---|---|
| Number of Observations with Available Data | 23 | 34 | 22 |
| Magnitude of Cost Overruns (Measured as Mean Overruns) | 159.30% | 0.00% | +1.10% |
| Frequency of Cost Overruns | 10 of 10 | <3.4 per 1 million (approx. 0 of 10) | 5 in 10*** |
| Variability of Cost Overruns (Measured as St. Dev. Of Overruns) | 160.32% | <10.00% | 10.00% |
| Radical Change in Delivery Format | 0 | >0 | >1 Shift from NASA's bespoke to a platform strategy |

*See Muth (1961).
**See Ansar and Flyvbjerg (2022).
***A good practical approximation for a portfolio manager.

Table 2. Cost overruns at the Olympics by Region (1960–2024), real terms.

| Region | Count | Frequency Cost Overrun % | Average Cost Overrun % | Variance (SD) Cost Overrun % |
|---|---|---|---|---|
| Asia-Pacific | 6 | 6 of 6 | 71.17 | 62.36 |
| Europe | 10 | 10 of 10 | 158.80 | 89.50 |
| North America | 6 | 6 of 6 | 216.17 | 272.32 |
| South America | 1 | 1 of 1 | 352.00 | NA |
| **Total** | **23** | **23 of 23** | **159.30** | **160.32** |

statistically significant differences between any regional pairs. Thus, all regions experience similarly poor performance, indicating no region performed better or worse than others over the aggregate period.

We analysed the Olympics cost overrun data (magnitude, frequency, variance) over time to assess trends. Figure 1 shows significant fluctuations in cost overruns for both Summer and Winter Games from 1968 to 2024, with high overruns persisting. The variance in overruns is higher for Summer Games (44,654.54; mean = 194.90) than Winter Games (11,761.91; mean = 131.92), indicating more inconsistency in Summer events. However, a Mann-Whitney U test found no significant difference in overruns between Summer and Winter Games ($U = 72.5$, $p = 0.664$), suggesting similar budget management challenges. These findings highlight ongoing difficulties in controlling Olympic costs, with no clear trend of improvement.

Following Montreal's 1976 debacle, which resulted in an extreme cost overrun of 720% in real terms, there is no Summer Olympics data available until another significant overrun in Barcelona in 1992, where the cost overrun was 266%. Although the trend line for the Summer Olympics appears to show a decline in cost overruns, the overall level remained elevated, and the variability in cost overruns for Summer Games also remained high, with significant fluctuations from one event to another.

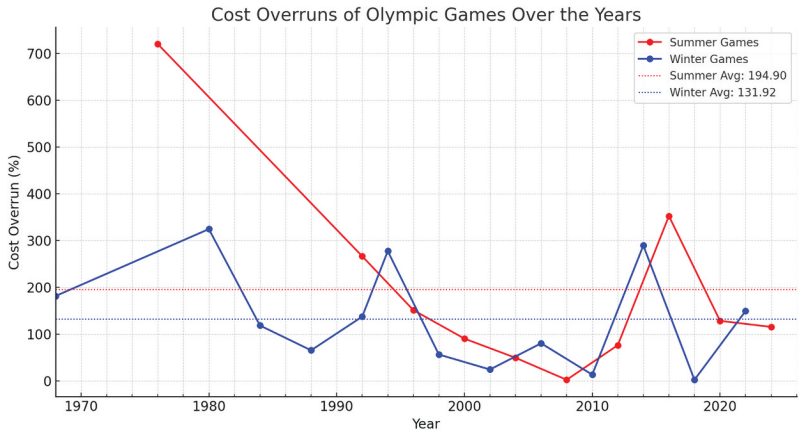

**Figure 1.** Cost overruns of Olympic Games over the Years, in real terms.

We conducted a comprehensive set of statistical tests to assess whether cost overruns in the Olympics show a progressive reduction over time. First, we compared linear and polynomial models. The linear model did not reveal a consistent downward or upward trend in cost overruns; however, residual patterns indicated mild curvature. Introducing a second-order polynomial (quadratic) improved the overall fit but did not fully satisfy normality assumptions. Despite these minor deviations, we deemed the quadratic model suitable for describing overall trends, as our primary goal was descriptive rather than reliant on strict inferential assumptions.

Using LOESS and polynomial models and treating the data as a time series with ARIMA(0,0,4), all analyses showed no sustained decline in Olympic cost overruns, only short-term fluctuations. Neither mean overruns nor variance have consistently decreased, and any recent dips are unlikely to be due to IOC initiatives such as the Knowledge Management Program or Agenda 2020. ARIMA forecasts over 12 years predict persistently high overruns, indicating no evidence of progressive learning.

Figure 2 depicts the ARIMA(0,0,4) forecasts for future Olympic cost overruns, highlighting the model's predicted trend alongside the range of likely values. The solid blue line shows the central forecast, the darker blue band represents the 80% prediction interval, and the lighter grey band represents the 95% prediction interval within which future outcomes are expected to fall. Under the current rotational hosting format, these large

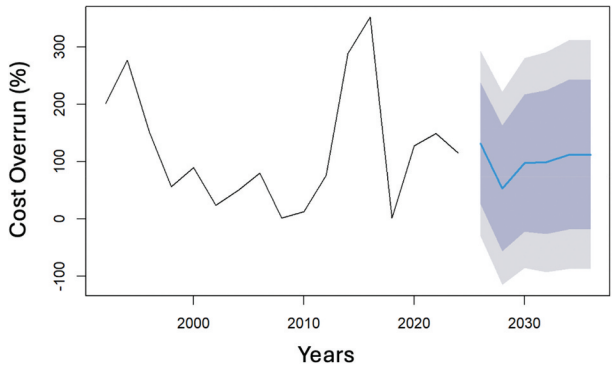

**Figure 2.** Predicted central tendency of cost overruns for future Olympics (2024 onwards), real terms.

overruns are likely to persist, underscoring the enduring challenges of managing Olympic costs.

To summarize the statistical analysis, taken as a whole, the Olympic Games have not demonstrated any reduction over time in cost errors, their magnitude, frequency, or variance. Host cities are no better at delivering the Olympics on budget today than they were 64 years ago. The IOC's Olympics Knowledge Management Program and Agenda 2020 have failed to make a lasting impact on the learning curve for host cities. Without a radical break from the past, predictive models suggest that the poor performance will remain difficult to escape. Such a pattern raises questions about the effectiveness of the learning mechanisms currently in place within the Olympic organizational structure.

These statistical patterns – persistent high mean overruns, universal frequency of overruns, and substantial variance – directly inform our central question: do the Olympics learn across space and time? The absence of improvement in these metrics over six decades supports our propositions about the 'myopia of learning' and sets up the discussion of why the Games fail to translate tactical adaptations into higher-level strategic learning.

While Olympic host cities struggle with cost overruns, other cities are learning by opting out of bids due to financial concerns. Rome's high-profile withdrawal from the 2020 and 2024 bids exemplifies this trend. In 2017, facing a lack of candidates, the IOC awarded the 2024 and 2028 Games without competitive bidding. Historically, the contest to host the Olympics has been fierce; the 2004 Summer Olympics attracted eleven bids, followed by ten for the 2008 Games and nine for the 2012 Games. However, recent years have witnessed a reversal of this trend, as knowledge of the financial pitfalls associated with the Olympics has become more pervasive. Recently, cities like Boston, Budapest, Munich, and Mexico City have retracted bids, citing financial risks (Bourbillères, Gasparini, and Koebel 2023; Fabry and Zeghni 2019; Kassens-Noor 2019; Lauermann 2022; Matheson and Zimbalist 2021; McBride and Manno 2024).

Paris 2024 has staked a bold claim to host a 'lean Olympics' with 95% of venues already in place and limited new construction (Soubrier, Mrsnik, and Tocanne 2024). Despite claims of frugality, a 115% cost overrun ensued. The IOC's Agenda 2020, aimed at cost control, has previously failed, as seen in Rio 2016, PyeongChang 2018, Tokyo 2020, and Beijing 2022. Fiscal restraints may suppress cost overruns temporarily, but extreme outcomes are inevitable in fat-tailed distributions, a phenomenon known as 'regression to the tail' (Flyvbjerg 2020; Flyvbjerg, Budzier, and Lunn 2021). Emerging market cities aspiring to host the Olympics face significant infrastructure challenges, making the 'lean Olympics' a rhetorical and unreliable strategy.

### Lower-level learning at the Olympics

'The next time they'll get it right'.

Cyclist Erika Salumäe (Estonia) and 1000 m gold medallist at Barcelona 1992 after the officials overseeing the medal ceremony hoisted the Estonian flag upside down (Witt 2012, 12).

The quantitative evidence highlighting the lack of meaningful higher-level learning by Olympics host cities presents a paradox. A burgeoning literature finds rich qualitative evidence of learning across the Games and other temporary organizations involved in hosting mega-events (see, for example, Bayle and Chappelet 2004 or Frawley and Daryl 2013). In Chatziefstathiou and Müller (2014), various contributing authors discuss the pedagogical and educational legacies of Baron Pierre de Coubertin and the Olympics. Similarly, Grabher and Thiel's (2015) study on the London 2012 Olympics offers a rich exploration of the learning trajectories in temporary organizations, particularly through the lens of project-based work. The authors illustrate how the Olympics served as a high-stakes, high-visibility project that facilitated a unique convergence of skilled professionals from various fields, fostering an intensive exchange of existing knowledge but also generating new insights through collaborative challenges (also see Thiel and Grabher 2015). Building on Müller and Stewart (2016), Wolfe (2023, 266) argues 'As mega-events are in more-or-less continual phases of bidding, preparation, and execution, organizers from different host cities routinely participate in knowledge transfer activities'. This literature emphasizes the deliberate effort to capture and share accumulated knowledge through 'learning legacy' initiatives. The learning highlights not only technical, logistical, and supply chain aspects but also tactics for stakeholder engagement and community involvement (Frawley and Daryl 2013; see also Werner, Dickson, and Hyde 2015 on the Rugby World Cup).

Specific examples of this transfer of knowledge include Improved crowd management in filling or emptying venues (Doukas 2006) including improved accessibility at venues through integrated wheelchair seating and tactile ground surface indicators (Darcy and Tracy 2013). During the 1980 Moscow Olympics, controversial fouls for scraping the swing leg on the ground in the Men's Triple Jump Finals, led to the IAAF re-evaluating and ultimately changing the rules (Junction Dir 2023). During the Rio 2016 Olympics, the transportation team learned to adapt their coordination efforts, relying heavily on informal mechanisms like real-time communication tools (Fernandes, Spring, and Tarafdar 2018). Tragic terrorist events at Munich 1972 and Atlanta 1996 led to extraordinary security measures to prevent another attack – albeit fat tail risk always remains (Clément 2017; Graham 2012; Michalski, Marek Radomyski, and Bernat 2023). A more sophisticated understanding of anti-doping is assisting with educating athletes and proactively identifying doping strategies (Hayward et al. 2022). Furthermore, the literature advocating that Olympics organizations are learning explores the implications of such temporary mega-events on professional communities and industries. The transient nature of the employment created by the Olympics, characterized by intense bursts of activity followed by rapid dissolution, create opportunities for how professionals learn from their Olympic experience to secure career advancement and consultancy roles that benefit from their enhanced reputations.

The qualitative literature paints a picture of dynamic, intense interactions of adapting the soaring, timeless, Olympic principles to local constraints bounded by space and time. This ad hoc mobilization of human energies is reminiscent of literature on surprise. Bechky and Okhuysen (2011), for example, compare ethnographic data from a police SWAT team and film product crews that routinely encounter surprising events requiring a swift response. The authors find that professional teams are surprisingly good at dealing with surprise – in their words, 'individuals engage in organizational bricolage, restructuring their activities by role shifting, reorganizing routines, and reassembling the work. Organizational bricolage depends on the socio-cognitive resources that group members develop by drafting agreement on the work, reinforcing and elaborating task activities, and building cross-member expertise' (Bechky and Okhuysen 2011, 239).

How do we reconcile quantitative evidence showing no learning with rich qualitative case study accounts of learning? The concept of organizational bricolage, although effective during unexpected moments, aligns with Fiol and Marjorie (1985) notion of lower-level learning. Perversely, the necessity for frequent micro-learnings (such as ensuring the flag of Estonia is correctly displayed) suggests a delivery process that is both fragile and prone to errors. Although mobilizing cross-member expertise is commendable and even heroic within Olympic delivery organizations, the ongoing need for such interventions also highlights a lack of consolidation of higher-level learning, contributing to the difficulty of hosting the Olympics within a reasonable budget and risk envelope. 'Little errors can point to bigger problems – just ask Van Halen', writes legal arbitrator Dr. Simpson (2021). Simpson (2021) recounts Van Halen's 'no brown M&Ms' clause as a test of whether venue staff had read critical technical requirements, its breach signalling deeper risks. Consistent with Proposition 2, reacting to hazards (e.g. preventing a fire; Venkataraman and Pinto 2017) is lower-level learning, whereas designing such checks into contracts reflects higher-level learning.

Revisiting our initial propositions, we now examine how our empirical findings confirm or illuminate each, underscoring that the spatial and temporal dispersion inherent in the Olympics hinders the systematic accumulation of learning from one Games to the next. *Proposition 1*. Insofar as we use cost estimation as our proxy for cross-Games learning, we find strong support for the view that organizational learning fails to arise spontaneously in an infrequent, globally dispersed setting like the Olympics. As the next section further illustrates, such learning only emerges when it is deliberately designed and embedded. See Bjerke-Busch and Thorp (2024) for corroboration from public management studies. *Proposition 2*. Our data show that while tactical improvements often occur within individual host committees, these lower-level learnings rarely accumulate into higher-level, strategic gains across successive Games. *Proposition 3*. Persistent budget overruns and operational vulnerabilities indicate that the cultural, administrative, geographic, and economic distances among hosts significantly impede effective inter-Games knowledge transfer.

In conclusion, both quantitative and qualitative evidence reveal a persistent pattern of learning failures within the Olympic framework. The current model, requiring a new host city every four years, effectively resets institutional knowledge, mirroring the myth of Sisyphus. This cyclical reinvention fuels recurring cost overruns and operational inefficiencies, as host cities struggle to integrate lessons from past Games. Despite the involvement of seasoned experts, the Olympic delivery process remains trapped in what Flyvbjerg, Budzier, and Lunn (2021) describe as the 'eternal beginner syndrome', where past mistakes are continuously repeated. Structural factors, such as the rotational hosting system, the incentive misalignment between the IOC and Local Organizing Committees (LOCs), and the vague guidance in official documentation, further inhibit sustained learning. Fragmented, ad hoc improvements at the project level have failed to translate into systemic reforms that prevent recurring inefficiencies. Given these fundamental shortcomings in knowledge transfer, the next section explores radical, research-backed strategies for transforming Olympic delivery to ensure long-term improvements.

## Discussion

Our point of departure was Levinthal and March's (1993) concept of 'myopia of learning', which suggests that learning within organizations is difficult, particularly across space, time, and failure. The Olympics exhibit all three. As established in our theoretical framework, the spatiotemporal dispersion of the Olympic Games creates structural barriers to higher-level learning. We now consider how these barriers might be addressed. In contrast to organizations like the IOC or FIFA that have struggled with repeated risk blowouts, empirical evidence suggests that certain organizations, such as Apple, Ferrari, Holiday Inn, IKEA, McDonald's, Nike, Nucor, Pixar, and Unilever have been more successful in managing spatiotemporal dispersion in their portfolios. These organizations have implemented various strategies to address the challenges associated with learning in projects spread across space and time. From the literature we infer four archetypal learning strategies – incremental, centralizing, decentralizing, and real options. Table 3 summarizes each of the strategies and provides an associated example. While these comparator firms are commercial and answerable to shareholders, the IOC and FIFA can still draw lessons from how large, dispersed corporations tackle recurring challenges – recognizing that their non-profit, membership-based models introduce added complexity but do not preclude adapting proven strategies. We now turn to how the Olympic Games may seek to implement one or more of these strategies. The strategies carry broader implications for public management theory, which we discuss later in the conclusion.

This contrast between companies such as Ferrari, Apple, Toyota, Merck, and Netflix and the Olympic ecosystem underscores that the IOC lacks various incremental, centralizing, decentralizing, and learning mechanisms, which, if present, could plausibly address the persistent cost overrun patterns identified in our results. In other words, if the IOC and the Olympic ecosystem were to continue using the current structure and rotational model, they cannot expect significant delivery improvements. In terms of our propositions, these strategies, which are not deployed by the Olympics ecosystem, are specifically designed by a permanent organization to counteract spatial and temporal constraints and induce higher-level learning.

### Incremental strategies

Incremental learning refers to continuous learning in stable environments, often described as 'White Swans', with relatively tight coupling. The literature on incremental learning presents a valuable yet lengthy list of potential initiatives, such as fostering a motivating vision, establishing clear accountabilities and monitoring systems, enforcing strong ethical boundaries, creating interactive learning opportunities, investing in personnel development, nurturing a supportive organizational culture, integrating technology with processes, and implementing repeatable formats and designs prior to achieving full modularity (see Ansar and Flyvbjerg 2022; Easterby-Smith and Lyles 2011; Senge [1990] 2006; Simons 1995). Companies like Holiday Inn, IKEA, Ferrari, McDonald's, and J.P. Morgan have reaped the benefits of such continuous learning practices (Jenkins and Tallman 2016; Jonsson and Foss 2011; Winter and Szulanski 2001).

**Table 3.** Four learning strategies.

| Learning Strategies | Archetypal Examples & Description | Reference |
| --- | --- | --- |
| 1. Incremental | **Ferrari**: From 1929 to 2008, Ferrari incrementally learned by incorporating global technical knowledge and architectural innovations, particularly from the UK's Motorsport Valley. This transitioned the company from a local, engine-centric focus in Italy to a global integration of advancements in chassis design and aerodynamics. | Jenkins and Tallman (2016) |
| 2. Centralizing | **Apple**: " … Apple's innovation is centralized in a single physical space, its campus in California, and you have an innovation hotbed. Buttress your market presence through clever, contrarian maverick branding, supported by radically new products, and you can develop a cult following". | Heracleous (2013, 95) |
| 3. Decentralizing | **Toyota**: Toyota's decentralized learning and supplier network leverage individual plants as mother plants, which utilize unique knowledge for global knowledge transfer, continuous improvement, and problem-solving, enhancing both manufacturing processes and supplier network efficiencies across international locations. | Suh (2019) |
| 4. Real Options | **Merck**: Pharmaceutical companies like Merck use real options reasoning to strategically manage R&D investments, treating initial efforts as options for future expansion. This approach allows them to invest sequentially, maintaining flexibility to adjust or abandon projects based on future learning and emerging outcomes, effectively balancing risk and potential rewards.<br>**Netflix**: Sequential, methodical, evolution from DVD rental, online streaming of other companies' content, to original content creation. | R. G. McGrath, Ferrier, and Mendelow (2004) Jaworski (2021) |

Despite their efforts, the Olympics have struggled to implement incremental learning strategies successfully. Unlike the tightly coupled and stable environments conducive to incremental learning, the Olympics are characterized by geographically dispersed events, temporal gaps, and fragmented local delivery organizations. The IOC has attempted to mitigate these challenges with guidelines, roving expert teams, and lofty rhetoric. However, evidence points to a consistently poor average performance. The promise of incremental learning has often been more about generating consulting revenues than achieving significant improvements in organizational outcomes. Following Levinthal and March's caution, we remain sceptical that incremental strategies alone can address the learning challenges at the Olympics. Instead, they must embrace radical approaches to overcome the 'myopia of learning'. Next, we review such radical approaches: centralizing, decentralizing, and real options.

### Centralizing strategies

A centralizing strategy could involve selecting a permanent location for the Olympic Games, one for the Summer and one for the Winter Games. This could significantly reduce risks associated with the continual need for new infrastructure by instead utilizing existing facilities for each iteration. Furthermore, these permanent locations would streamline organizational processes and enhance economic predictability. They would also facilitate the accumulation of expertise, enabling proactive measures to mitigate security risks and potential environmental impacts, thereby eliminating many of the spatial constraints on learning and adaptation.

The concept of a permanent Olympic venue has a distinguished historical precedence. For nearly twelve centuries – from 776 BC to AD 393 – the ancient Olympic Games were held in Olympia, located in the western Peloponnese peninsula of Greece. These quadrennial games drew competitors and spectators from across the Greek world. Reflecting on this tradition, in 1896, King George I of Greece proposed Athens, the cradle of the ancient Olympics, as the permanent site for the modern games, arguing that it would bolster Greece's tourism-reliant economy. The idea was later supported by Pierre de Coubertin in the early 1900s, suggesting a deep-rooted value in stability (Silvestre et al. 2024). Athens would thus form a natural candidate for such a permanent home. Los Angeles, with its existing facilities and ongoing upgrades to public transportation in preparation for the 2028 Olympics, also stands out as a strong contender for the Summer Games, particularly given its profitable experience of hosting the 1984 Olympics. For the Winter Games, a location with a reliable natural snow supply, despite the challenges of global climate change, would be ideal, assuming sufficient public backing.

The centralization strategy does entail drawbacks, such as the discontinuation of the Olympics' global rotation and increased local disruptions in the selected location. Despite these issues, the benefits of reducing significant risks and enhancing overall alignment with both ancient traditions and contemporary needs likely outweigh the disadvantages. A variation of the centralizing strategy could be to elect a rotating model among a few cities that would preserve the international spirit of the games while reducing the exorbitant costs associated with constant infrastructure building and bidding.

### *Decentralizing strategies*

The decentralizing strategies propose decomposing the Olympics into their constituent events and assigning each a permanent home globally. This approach, segueing from the centralizing strategy, involves establishing event or venue-level permanent homes that promote reuse and significantly mitigate risks associated with cost overruns from constructing new venues each time. By decomposing and distributing the events worldwide, this strategy not only maintains the Olympic international spirit but also capitalizes on the benefit of spatial accumulation of learning by event type.

There is some historical precedent for decentralization in that the IOC has increasingly allowed Olympic host cities to spread events across neighbouring towns, regions, and even other countries. For example, Paris 2024 hosted the surfing competition in Teahupo'o, Tahiti, one of the world's most renowned surfing spots, marking one of the longest distances between the main host city and an event venue in Olympic history. Similarly, previous Games have utilized strategic locations to maximize existing facilities and distribute economic benefits – London 2012's sailing events took place in Weymouth, Beijing 2008's equestrian events were in Hong Kong, and Pyeongchang 2018 held ice events in Gangneung. Winter Olympics like Vancouver 2010 and Salt Lake City 2002 also capitalized on regional venues such as Whistler and Snowbasin to meet specific sports requirements.

A more radical extension of decentralization strategies might also involve dissolving the traditional quadrennial format, allowing Olympic events to occur asynchronously and be marked on the sporting calendar as Olympic events whenever and wherever they take place. This innovative approach could enhance learning not only through the spatial specialization by event but also through more frequent temporal occurrences, potentially fostering clustering or agglomeration effects around specific event types. For example, certain host locations could develop an economic base centred around their chosen event. Skiing 'World Cup' events already follow such a format.

Decentralizing strategies could, however, present logistical challenges for athletes dispersed by event type across multiple locations, complicating travel and participation for those competing in multiple events. Additionally, this approach may lead to a more complex viewing schedule for audiences, spanning multiple time zones. Despite these challenges, the potential benefits of fostering event-based expertise and encouraging economic specialization in each location could outweigh these drawbacks, promoting sustainability and economic growth within host communities, and as such are worth considering.

### *Real options strategies*

Real options are strategic decisions that span time, offering flexibility and the potential to adjust course based on future developments. They offer opportunities to learn as executives make sequential investments. They provide decision-makers with the ability to adapt and respond to uncertainty, thereby enhancing the value and strategic positioning of the firm (see G. McGrath, Ferrier, and Mendelow 2004 for a literature review and exposition; also see and R. G. McGrath 1997).

Real options strategies, when applied to the Olympics, would help host cities construct sporting venues suitable for their current level of development and gradually enhance these facilities as demand becomes clearer. Consider Karachi as an example: presently, it is not primed to host the Olympics, but it boasts a large local population that could use new sporting facilities, albeit tailored to a modest scale befitting Pakistan's current GDP. As Karachi evolves and the use of the sporting infrastructure becomes more established, the expansion could progress to accommodate various events sequentially. Initially, the focus may be on local events like university competitions, followed by national gatherings, then regional games such as the South Asian Games, and eventually, international showcases like the Commonwealth Games and even the Olympics.

The real options approach can be integrated with either a centralized strategy, cultivating a rotation of host cities, or a decentralized approach, establishing permanent venues based on event type. The fundamental notion is that constructing Olympic-grade facilities requires time and ensuring their ongoing utility demands even more. *Ex ante*, it's uncertain whether a specific location will emerge as an attractive event host. For instance, following the 1980 Winter Olympics, Lake Placid's Olympic village was repurposed into the Federal Correctional Institution Ray Brook merely six months later by the U.S. Bureau of Prisons due to uncertainty surrounding its appeal to winter sports tourists. By harnessing real optionality across multiple cities, a broader pool of potential host candidates can evolve gradually. This approach enables cities to benefit from increasingly improved sporting facilities tailored to their local population, while also spreading costs over an extended timeframe in alignment with improved demand and usage data.

Real option strategies systematically exploit time as a facilitator of learning. R. G. McGrath, MacMillan, and Venkataraman (1995) examine organizational competence, defining it as the ability of a firm or its units to consistently meet or surpass goals. They highlight two important factors, 'comprehension' (understanding tasks) and 'deftness' (skill in task execution), as essential for competence. Through a study covering 160 initiatives across 40 organizations in 16 countries, they find that comprehension and deftness can be developed over time. This process-oriented paradigm proposes an iterative sequence: probing, learning, revising, consolidating, scaling. Real options thus provide a strategic framework for systematic development of competence over time.

The real options strategies for the Games may face drawbacks such as difficulty in accounting for the true infrastructure costs, uncertain demand projections, or loss of momentum and drift. However, these drawbacks can be mitigated through a clear orientation and adaptive decision-making processes that respond to evolving circumstances and market conditions aligned closely to local sporting needs ahead of international competitions.

### Implications for public management

Organizational learning within public management is a critical yet under-researched area that requires further scholarly attention to improve the effectiveness of public institutions (Rashman, Withers, and Hartley 2009). The Olympics, as a case study of public organizations, demonstrate significant challenges in achieving higher-level learning, reflecting broader issues with learning in public administration. Our empirical critique aims not to diminish the efforts of those involved in the delivery of challenging public endeavours but to highlight systemic political biases that overlook critical information, such as cost overruns, thereby affecting the transfer or learning (Boyne 2003; March and Olsen 1983). Higher-level learning is crucial for delivering public value, yet numerous barriers within the public sector hinder this process (Rashman, Withers, and Hartley 2009).

David A. Good (2007) elucidates that even in a democracy like Canada, the federal budgeting process involves intricate and opaque interactions among spenders, guardians, priority setters, and financial watchdogs. These evolving roles and relationships significantly impact the effectiveness and transparency of fiscal governance. Good's recommendations, including the creation of an expenditure review committee of cabinet and reassigning expenditure review tasks to the Treasury Board Secretariat, aim to improve the budgeting process through higher-level learning. The barriers to high-level learning aligns with Levinthal & March's concept of the myopia of failure, where there is a tendency to ignore uncomfortable information about failures. Rayner (2012) terms this 'uncomfortable knowledge'. Public administration appears particularly prone to this phenomenon. Radical candour – caring deeply about public value and confronting challenging evidence directly – is an essential foundation for such higher-level learning in public administration.

Incrementalism is a prevalent characteristic in public administration due to the sector's stability and longevity. This approach is often justified as the public sector navigates complex problems without fixed solutions (Lindblom 1959; Osborne and Brown 2005). Modern governance requires robust mechanisms for accountability and performance management, which can make governmental processes slow and deliberative (Kettl 2015). However, public administration should adopt a bolder approach and be willing to abandon failed actions once sufficient evidence mounts against them. Moynihan (2005, 2008) highlights the importance of goal-based learning and dynamic performance management, which necessitates responsiveness and adaptability. Hood and Margetts (2007) discuss the evolution of governance tools, emphasizing the need for modern public administration to be agile and evidence driven. As argued above, the rotational model of the Olympics is flawed and should be replaced with a radically different approach from the alternatives we highlighted. This transformation does not require abandoning the Olympics' ideals of international spirit or athletic achievement. Similarly, in public management, innovative strategies should be identified to make delivery more affordable and effective without diluting the higher societal purpose.

Understanding the spatial and temporal dimensions in public administration is crucial for managing dispersed projects and ensuring the success of complex programmes like the Olympics. In federated countries such as the US, India, Australia, Canada, and Brazil, policy solutions effective in one state may not be directly transferable to another. Additionally, the temporal nature of policy initiatives means that past successes might not be applicable in the present, requiring continuous, critical assessment. Incorporating global learning systematically can address this myopia. Our study underscores the necessity of evaluating the long-term impacts and effectiveness of public projects. As highlighted by Wond and Macaulay (2011), longitudinal research can provide extensive contextual detail that enhances public management evaluation, allowing for a more comprehensive understanding of project impacts over time. The impact of spatial and temporal considerations is also relevant in diverse contexts such as university administration (e.g. the autonomous operation of colleges within Oxford or Cambridge), multinational utilities, and public management with centre-periphery tensions. Effective levers of learning must be identified and utilized to ensure alignment and prevent overruns (Hartley 2008; Pettigrew 2005).

The concept of legacy in public management, referring to the long-term impact and sustainability of large-scale projects, is vital. As noted by Gillett and Tennent (2022), the legacy of mega-events like the Olympics is influenced by complex institutional goals and the necessity to balance various stakeholders' interests. The Olympics' failure to achieve strategic learning and control costs threatens their legacy, making future host cities wary. This issue extends to other public sector organizations like the United Nations and the World Bank. Public management theory must emphasize strategic approaches to ensure the long-term sustainability and success of large-scale projects (Moore 1995, 2014; Osborne and Brown 2005). We suggest parsimony in the liberal use of the word, 'legacy'. 'Learning legacies' such as those at the London Olympics, often celebrated within government, are at risk of becoming mere political rhetoric without substantial impact. These collections frequently consist of fragmented, tactical learnings. It is imperative for the public sector to strive for more substantial and higher-level learning initiatives (Tomlinson 2014).

In sum, the new literature discussed here provides theoretical and comparative support for interpreting the empirical findings of persistent cost overruns and high variance as evidence of a myopia of learning in the Olympic system.

## Conclusion

The persistent challenges in managing the Olympics underscore a fundamental problem in how projects learn across space and time. Despite the repetitive structure of the Games, the spatiotemporal dispersion inherent in the rotational hosting format, compounded by the temporary organizational form of each host committee, blocks the accumulation of higher-level learning. Our statistical analysis confirms this: over 64 years and 23 Games, there has been no sustained reduction in the magnitude, frequency, or variance of cost overruns. Tactical learning abounds within individual Games, but none of it aggregates into the strategic improvements needed to contain costs or transform delivery.

Building on Levinthal and March's (1993) myopia of learning, we developed the concept of spatiotemporality to explain this pattern: geographic distance makes knowledge context-dependent, temporal distance ensures it degrades before the next iteration, and the temporary organizational form of each OCOG ensures that the mitigating structures available to permanent organizations are absent by design. The Olympic franchise model, in which a permanent franchisor delegates delivery to a rotating series of temporary franchisees, represents an extreme case of this compound challenge, but it is not a unique one. Public institutions managing large-scale programmes across dispersed geographies and extended time horizons, from peacekeeping operations to federated infrastructure systems, face structurally analogous barriers to learning.

To overcome the spatiotemporal barrier, we proposed four learning strategies: incremental, centralizing, decentralizing, and real options. The Games have relied almost exclusively on incremental approaches. We have shown this to be insufficient. More radical structural reform, whether centralizing the Games in permanent locations, decentralizing events across specialized global venues, or adopting real options for sequential capability building, offers a more credible path to sustained learning and cost containment.

This study contributes to public management theory by introducing spatiotemporality as a framework for understanding learning failures in dispersed project ecologies, extending the myopia of learning into the spatial and temporal domains. It contributes to practice by demonstrating that without deliberate structural intervention, the Olympic model will continue to produce the outcomes it has always produced. The challenge for the IOC and future host cities is whether they are willing to confront this evidence or continue to treat each new Games as though no one has ever staged one before.

## Notes

1. Projects are profound human attempts to overcome the constraints of space and time. With the earliest dams, built nearly 5,000 years ago, humans stored and transported water to overcome the limitations of where and when water would be available. Roads followed suit. Spaceflight now explicitly seeks to build a multi-planet human society. Flyvbjerg et al. write (2003, 19), 'Megaprojects form part of a remarkably coherent story. Sociologist Zygmunt Bauman perceptively calls it the "Great War of Independence from Space", and he sees the resulting new mobility as the most powerful and most coveted stratifying factor in contemporary society. Paul Virilio speaks of the "end of geography" while others talk of the "death of distance"'. Although projects are instruments for human independence from space and time, they themselves are constrained by the vagaries of spatial and temporal issues.
2. Large multinational organizations, notwithstanding their seemingly monolithic and tightly controlled façade, similarly grapple with learning across dispersed business units and projects (Söderlund 2004). The irony is sharp: the very organizational forms designed to conquer distance are themselves undone by it.
3. The IOC has institutionalized knowledge transfer through its Olympic Games Knowledge Management programme, combining formal tools (e.g. technical manuals, debriefs) with informal mechanisms (e.g. staff secondments) (Qin, Rocha, and Morrow 2022). Recent governance reforms have sought to improve transparency, accountability, and stakeholder inclusion in response to legitimacy challenges (Bayle 2024; also see; Chappelet 2008, 2023). Whether these mechanisms are adequate is precisely the question our cost data address.
4. The bidding cycle itself constrains learning: because host cities only confirm their status after securing the Games, the time available to systematically acquire and adapt lessons from previous hosts is compressed, often committing them to sub-optimal knowledge search and transfer.
5. 山高皇帝远 - Shan gao, huangdi yuan illustrates the timeless phenomenon of control in a geographically dispersed setting.
6. Catastrophic forgetfulness, also known as catastrophic interference or catastrophic forgetting, is a problem encountered in neural networks and artificial intelligence systems. It occurs when a neural network, after learning a new task or dataset, forgets or interferes with the previously learned information. This phenomenon is particularly relevant to connectionist models and deep learning architectures, which are designed to learn incrementally and adaptively.

## Acknowledgments

We are grateful to Mariagrazia Zottoli of the Oxford Statistical Consultancy Unit (OxSC), Department of Statistics, University of Oxford, for her expert guidance on the statistical analyses presented in this paper. We also thank the editors and anonymous reviewers at *Public Management Review* for their constructive and generous engagement with our work. Their comments substantially improved the manuscript.

## Disclosure statement

No potential conflict of interest was reported by the author(s).

## Funding

This work was supported by the Smith School of Enterprise and the Environment at the University of Oxford through a research grant from Marex Group plc. Marex was not involved in the research design, methodology, data collection, analysis, or preparation of this article.

# Appendix. Olympic Games cost overruns

Table A1. Olympics dataset, cost overrun in real terms.

| Year | Cost overrun % | Olympic Games | Type | Region |
| --- | --- | --- | --- | --- |
| 1968 | 181 | Grenoble | Winter | Europe |
| 1976 | 720 | Montreal | Summer | North America |
| 1980 | 324 | Lake Placid | Winter | North America |
| 1984 | 118 | Sarajevo | Winter | Europe |
| 1988 | 65 | Calgary | Winter | North America |
| 1992 | 266 | Barcelona | Summer | Europe |
| 1992 | 137 | Albertville | Winter | Europe |
| 1994 | 277 | Lillehammer | Winter | Europe |
| 1996 | 151 | Atlanta | Summer | North America |
| 1998 | 56 | Nagano | Winter | Asia-Pacific |
| 2000 | 90 | Sydney | Summer | Asia-Pacific |
| 2002 | 24 | Salt Lake City | Winter | North America |
| 2004 | 49 | Athens | Summer | Europe |
| 2006 | 80 | Torino | Winter | Europe |
| 2008 | 2 | Beijing | Summer | Asia |
| 2010 | 13 | Vancouver | Winter | North America |
| 2012 | 76 | London | Summer | Europe |
| 2014 | 289 | Sochi | Winter | Europe |
| 2016 | 352 | Rio | Summer | South America |
| 2018 | 2 | PyeonChang | Winter | Asia |
| 2020 | 128 | Tokyo | Summer | Asia |
| 2022 | 149 | Beijing | Winter | Asia |
| 2024 | 115* | Paris | Summer | Europe |

*At the time of writing (May 2024), Paris 2024 had not yet been delivered. Unlike the other numbers in the table, the cost overrun for Paris 2024 is therefore based on an estimate. Future overrun, calculated on the basis of actual outturn costs after the Games have been delivered, could therefore be higher than the 115% listed here.